\documentstyle[12pt,epsfig]{article}  
  
\newlength{\dinwidth}  
\newlength{\dinmargin}  
\def\lapproxeq{\lower .7ex\hbox{$\;\stackrel{\textstyle  
<}{\sim}\;$}}  
\def\gapproxeq{\lower .7ex\hbox{$\;\stackrel{\textstyle  
>}{\sim}\;$}}  
\def\be{\begin{equation}}  
\def\ee{\end{equation}}  
\def\bea{\begin{eqnarray}}  
\def\eea{\end{eqnarray}}  
  
\def\funp{{I\!\!P}}

\begin{document}  
\begin{flushright}  
DTP/00/40 \\  
15 September 2000 \\  
\end{flushright}  
  
\vspace*{2cm}  
  
\begin{center}  
{\Large \bf The compatibility of diffractive hard scattering} \\  
  
\vspace*{0.5cm}  
{\Large \bf in $p\bar{p}$ and $ep$ collisions}  
  
\vspace*{1cm}  
V.A. Khoze$^{a,b}$, A.D. Martin$^a$ and M.G. Ryskin$^{a,c}$  \\  
  
\vspace*{0.5cm}  
$^a$ Department of Physics, University of Durham, Durham. DH1 3LE \\  
$^b$ Theory Division, CERN, Geneva 23, Switzerland \\
$^c$ Petersburg Nuclear Physics Institute, Gatchina, St.~Petersburg, 188350, Russia  
\end{center}  
  
\vspace*{1cm}  
  
\begin{abstract}  
We show that the data for diffractive dijet production at the Tevatron $p\bar{p}$ collider are  
consistent with diffractive deep inelastic $ep$ data collected at HERA.  The breakdown of  
factorization between the two data sets is naturally explained by a strong tendency to populate  
the rapidity gap in the $p\bar{p}$ diffractive process by secondaries from soft underlying  
interactions and by bremsstrahlung associated with the presence of the hard dijet subprocess.  
\end{abstract}  
  
\vspace*{0.5cm}  
  
Nearly 40 years ago it was predicted that hadronic total cross sections would approach a  
constant asymptotic limit.  The Regge trajectory whose exchange ensures this behaviour  
became known as the Pomeron, with intercept $\alpha (0) = 1$.  Even today the observed  
slowly rising (high energy) total cross sections, and elastic scattering behaviour in the near  
forward direction, are remarkably well described by an effective trajectory $\alpha (t) \simeq  
1.08 + 0.25 t$, where $t$ is the square of the 4-momentum transfer (in units of GeV$^2$).  
Nowadays this is known as the \lq\lq soft\rq\rq\ or \lq\lq non-perturbative\rq\rq\ Pomeron.  
  
More recently, interest in Pomeron physics has been revived by studies of \lq\lq diffractive  
events\rq\rq\ in proton-proton collisions, which contain a rapidity gap in the final state, such  
that the hadrons produced in the collision only populate part of the detector away from the  
direction of one of the outgoing protons.  A supplementary condition for the presence of soft  
Pomeron exchange is that there should be a slow variation of the cross section as a function  
of the width of the rapidity gap.  The recent interest dates from the Ingelman and Schlein  
conjecture \cite{IS} that there would also be \lq\lq hard\rq\rq\ diffractive events in which the  
final state contains jets as well as a rapidity gap, and that the Pomeron (associated with the  
gap) is treated as a real particle made up of quarks and gluons which take part in the hard  
subprocess.  Such \lq\lq hard\rq\rq\ diffractive events have indeed been observed in high  
energy $p\bar{p}$ collisions, originally in the UA8 experiment \cite{UA8} at CERN, and  
most recently by the CDF collaboration  \cite{CDF} at the Tevatron $p\bar{p}$ collider.  The  
CDF collaboration study dijet production (for jets with $E_T > 7$~GeV) in diffractive events  
with a leading antiproton with a beam momentum fraction $x_F$ in the interval $0.905 < x_F  
< 0.965$, at $\sqrt{s} = 1800$~GeV.  For a given $x_F$ the rapidity gap is $\Delta y \sim \ln  
(1/(1 - x_F))$.  
  
Similar hard diffractive events are seen in high energy deep inelastic electron-proton  
collisions in which the outgoing proton travels approximately in the original beam direction  
leaving a large gap between its rapidity and that of the other hadrons \cite{H1,ZEUS}.  We  
speak of diffractive deep inelastic scattering (DDIS).  The DDIS cross section can be  
factorized \cite{COL} into a convolution of \lq\lq universal\rq\rq\ parton densities of the  
Pomeron (sometimes called diffractive parton distributions) with the partonic-level cross  
sections of the hard subprocess, see Fig.~1(a).  This is in direct analogy with the parton model  
of ordinary DIS in which we measure the universal parton densities of the proton.  Thus the  
HERA DDIS data may be used to constrain the parton densities of the Pomeron, see, for  
example, Ref.~\cite{H1}.  

\begin{figure}[t]
\begin{center}
\epsfig{figure=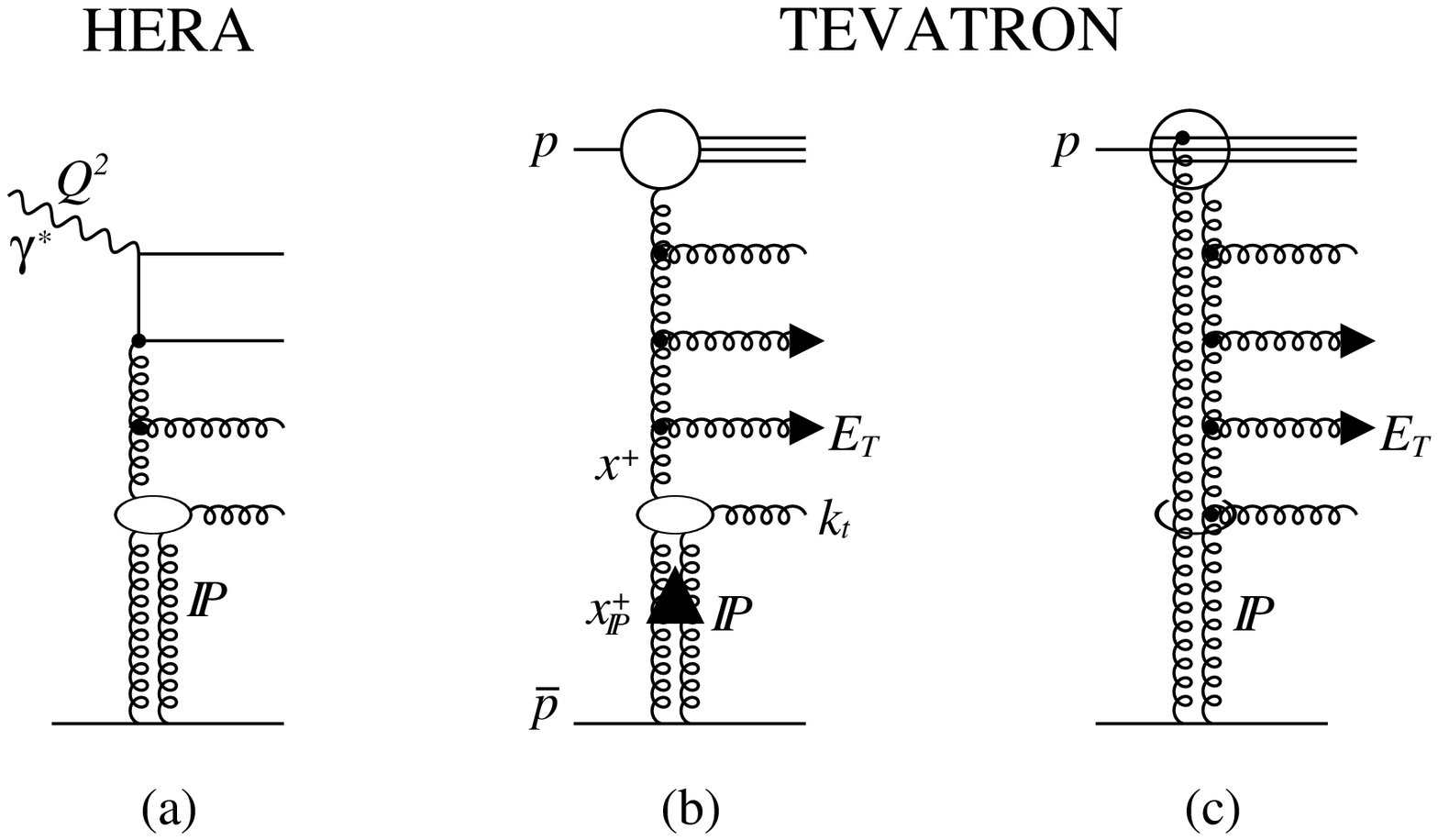,width=16cm}
\end{center}
\caption{
Diagrammatic representation of (a) diffractive DIS, (b) diffractive
dijet production in $p\bar{p}$ collisions, and (c) an additional,
non-factorizing, dijet production mechanism where the colour screening
gluon couples to a spectator parton.  In diagrams (a) and (b) the
Pomeron is treated as a real particle and the process is mediated by a
gluon input structure function of the Pomeron at scale $k_t$.  $E_T$
denotes the transverse energy of the hard jets emitted in the
diffractive $p\bar{p}$ process.
}
\end{figure}

However it is found \cite{ACTW,CDF} that when, as in Fig.~1(b), the parton densities of the  
Pomeron are used, together with the parton densities of the proton, to estimate the cross  
section for the hard diffractive dijet production observed in $p\bar{p}$ collisions, the  
factorized prediction turns out to be an order of magnitude larger than the data \cite{CDF}.  
To be precise the average \lq\lq discrepancy\rq\rq\ in normalisation is  
\be  
\label{eq:a1}  
D \; \equiv \; \frac{\rm data}{\rm prediction} \; \simeq \; 0.06 \: \pm \: 0.02.  
\ee  
A similar \lq\lq discrepancy\rq\rq,  
\be  
\label{eq:a2}  
D_W \; \simeq \; 0.18 \: \pm \: 0.04,  
\ee  
was found earlier between the predictions and the observations of diffractive $W$ production  
in $p\bar{p}$ collisions \cite{W}.  
  
A key assumption of the above factorization estimates is that the survival of the rapidity gap  
(associated with Pomeron, $\funp$, exchange) is the same in Figs.~1(a) and 1(b).  Here we  
emphasize that a breakdown of factorization is an evident consequence of QCD and occurs  
naturally due to the small probability,  
\be  
\label{eq:a3}  
\omega \; = \; S^2 T^2,  
\ee  
for the rapidity gap to survive in hadron-hadron collisions.  First, there is the probability, $1 -  
S^2$, that the gap in Fig.~1(b) may be filled by secondaries produced (via parton rescattering)  
in the underlying soft interaction; note that there is no such rescattering in DIS of Fig.~1(a).  
Second, there is the probability, $1 - T^2$, that the gap will be populated by extra gluon  
emission associated with the presence of the hard subprocess in diffractive hadron-hadron  
collisions.  A similar form (\ref{eq:a3}) of the survival probability  
was computed in Ref.~\cite{KMR} for the central production of a Higgs boson (and also of a  
dijet system) with a rapidity gap on either side.  Of course the survival probability $\omega$  
depends on the particular process, on the incoming energy, and on the final state  
configuration.  
  
The probability not to have extra soft rescattering has been estimated to be $S^2 \sim 0.01-  
0.1$ for various processes at different collider energies \cite{KMR}--\cite{KMR2}.  We refer  
to Ref.~\cite{KMR} for our most recent calculation of $S^2$, using explicit models for soft  
rescattering; see also \cite{GLM2}.  There it was found that the value of $S^2$ depends  
mainly on the optical density (or opacity) of the $p\bar{p}$ interaction as a function of the  
impact parameter $\rho_T$.  Briefly, the survival probability $S^2$ was calculated from  
\be  
\label{eq:b3}  
S \; = \; \langle \exp (- \Omega (\rho_T)/2) \rangle  
\ee  
where the average was taken over the $\rho_T$ dependence, and the opacity was assumed to  
have the Gaussian form  
\be  
\label{eq:c3}  
\Omega (\rho_T) \; = \; \frac{C^2 \sigma_0 (s/s_0)^\Delta}{2 \pi B} \: \exp  
(- \rho_T^2/2B), \ee where the slope of Pomeron exchange amplitude \be  
\label{eq:d3}  
\textstyle{\frac{1}{2}} B \; = \; B_0 \: + \: \alpha_\funp^\prime \: \ln  
(s/s_0),   
\ee  
and $C$ specifies the amount of anti(proton) dissociation in the rescattering  
process.  The parameters $\sigma_0, B_0, \Delta$ and $\alpha_\funp^\prime$  
of the eikonal model were tuned to describe the behaviour of the total and elastic differential  
$pp$ (or $p\bar{p}$) cross sections throughout the ISR to Tevatron energy range $(30 <  
\sqrt{s} < 1800~{\rm GeV})$.  Taking the mean value of $\rho_T$ to be the proton radius, it  
was found \cite{KMR} that $S^2 \sim 0.1$ at the Tevatron energy (and up to  
an order of magnitude smaller still at LHC energies).  
  
The factor $S^2 \sim 0.1$ in (\ref{eq:a3}) explains the main part of the \lq\lq  
discrepancy\rq\rq\ reported in Ref.~\cite{CDF}.  Some indication in favour of a small  
survival probability $(S^2 \lapproxeq 0.1)$ has also been observed by the D0 collaboration  
\cite{D0} in the process with two large $E_T$ jets separated by a rapidity gap.  Furthermore,  
there is a plausible explanation why $D < D_W$, as seen in (\ref{eq:a1}) and (\ref{eq:a2}).  
The optical density is smaller in the periphery of the proton.  On the other hand there are  
indications that the radius of the spatial distribution of quarks in the proton is larger  
than that for gluons\footnote{A comparison of the slope of diffractive $J/\psi$
photoproduction ($b\simeq 4~{\rm GeV}^{-2}$) with the behaviour of the proton electromagnetic
form factor ($b\simeq 5.5~{\rm GeV}^{-2}$), indicates that gluons have a smaller spatial
extent than quarks.  A similar conclusion follows from a study of QCD sum rules
\cite{BRAUN}}.  Since the $W$ boson is produced dominantly by quarks, whilst high  
$E_T$ dijets originate mainly from gluons, we expect the survival probabilities of diffractive  
production to satisfy  
\be  
\label{eq:a4}  
S^2 (W) \; > \; S^2 ({\rm dijet}).  
\ee  
  
Using (\ref{eq:b3})--(\ref{eq:d3}) we may estimate how $S^2$ depends on the size of the   
rapidity gap.  For the triple Pomeron process shown in Fig.~1(b)  
\be  
\label{eq:b4}  
\langle \rho_T^2 \rangle \; = \; b_0 \: - \: 2 \alpha_\funp^\prime \ln (s/M^2),  
\ee  
where $M$ is the invariant mass of the outgoing state produced by the $\funp p$ system.    
Thus, for example, as $x_F \rightarrow 1$ and the size of the gap increases (that is $s/M^2 =   
1/(1 - x_F)$ increases), slightly smaller values of $\rho_T$ are sampled and this, in turn, 
gives a little smaller $S^2$.  However this is not a strong effect, since $\langle \rho_T^2 
\rangle$ is dominated by $b_0$ (which is independent of $s/M^2$ and determined mainly by 
$B_0$).  
  
On the other hand the data show some $\beta$ dependence, where $\beta$ is the  
momentum fraction of the Pomeron entering the hard subprocess, $\beta = x^+/x_\funp^+$ of  
Fig.~1(b).  The CDF collaboration plot the discrepancy $D$ as a function of $\beta$ (see 
Fig.~4 of \cite{CDF}). They observe that $D$ decreases with increasing $\beta$.  The $\beta$  
dependence can be attributed to the behaviour of the survival factor $T^2$ of the rapidity gap  
against bremsstrahlung associated with the presence of the hard subprocess in Fig.~1(b).  The  
result depends on the scale $k_t$ at which the gluon structure function of the Pomeron is  
sampled.  Note that in Fig.~1(b) the lowest emitted gluon along the chain has transverse  
momentum close to $k_t$, whereas the hard jets have transverse energy $E_T$.  
  
The bremsstrahlung into the rapidity gap can originate from either the hard  
subprocess or associated with the Pomeron.  First, consider bremsstrahlung from  
 one of the hard $E_T$ jets.  When $\beta \rightarrow 0$, the hard jets are far  
 from the rapidity gap and so there is no radiation into the gap and no  
suppression of the diffractive process.  On the other hand, when $\beta  
\rightarrow 1$ one of the large $E_T$ jets becomes the lowest gluon in  
Fig.~1(b), $k_t$, and bremsstrahlung will populate the gap, which is now adjacent in phase  
space.  In this limit, if $E_T$ is sufficiently large, the suppression is, in principle, calculable  
from perturbative QCD, but will be sensitive to the experimental jet-searching algorithm.  
  
Now consider the $\beta \rightarrow 0$ and $\beta \rightarrow 1$ limits for bremsstrahlung  
associated with the Pomeron.  For $\beta \rightarrow 0$, the scale $k_t^2$ of the Pomeron   
structure function is soft.  There may be emission into the gap, but it is not calculable  
perturbatively.  However, in this limit, it is the same Pomeron, with the structure function,   
and the same emission, as that measured in diffractive DIS at HERA, and so there is no {\it   
extra} suppression in the diffractive $p\bar{p}$ process.  The situation is quite different for   
$\beta \rightarrow 1$, for which the scale $k_t \sim E_T$.  The configuration of the two   
$t$-channel gluons forming the Pomeron is now asymmetric and we have more emission into  
the gap than in DDIS\footnote{For DDIS at scale $Q^2\sim E^2_T$, it is   
possible to have Pomeron configurations with a large $k_t$ quark, but for gluons in DDIS we   
still have $k_t\ll E_T$.}.  In fact for sufficiently large $E_T$ ($\sim$ 50~GeV), the   
suppression of the diffractive $\bar{p}p$ process, arising from emission from the asymmetric  
gluon configuration of the Pomeron, can be estimated from perturbative QCD (see  
\cite{KMR}).  
  
In conclusion we have a qualitative understanding of the $\beta$ dependence.  For $\beta  
\rightarrow 0$ we expect (in comparison to DDIS) little extra suppression of the diffractive  
$p\bar{p}$ process from {\it bremsstrahlung} either from the hard subprocess or from the  
Pomeron.  That is $T^2 \simeq 1$ in this limit.  For increasing $\beta$, the suppression due to   
radiation increases (and $D(\beta)$ decreases) and, in fact, becomes calculable as $\beta   
\rightarrow 1$ if $E_T$ is sufficiently large (with the Pomeron structure function providing  
an effective infrared cut-off via a factor $(1 - \beta)^n$, where $n (k_t, E_T)$ is   
perturbatively calculable).  Although the discussion of the $\beta$ dependence has necessarily   
been qualitative, it is encouraging that the main trend is clearly seen in the data, see Fig.~4 of   
\cite{CDF}.  In fact we may identify $D (\beta) \sim 0.1-0.15$ for $\beta \lapproxeq 0.1$ with   
the survival factor $S^2$, since for $\beta \rightarrow 0$ we expect $T^2 \approx 1$.  At   
present it is not possible to make a reliable comparison of diffractive $p\bar{p}$ and DDIS   
data much below $\beta \sim 0.1$, since there are no DDIS measurements in this region.  
  
Although the survival factor $S^2$ is responsible for the main breakdown of  
factorization, we note that there is another non-factorizing contribution in diffractive  
$p\bar{p}$ collisions\footnote{It was noted in \cite{COL,CFS} that factorization is not valid  
in $p\bar{p}$ collisions.  A discussion can be found, for example, in the review of  
Ref.~\cite{WM}.}.  Besides the graph of Fig.~1(b) in which the second $t$-channel  
gluon (which screens the colour flow in rapidity gap interval) couples to a parton in the  
Pomeron fragmentation region (near the gap edge), there is also the possibility that the  
screening gluon couples to a fast spectator in the proton fragmentation region, see Fig.~1(c).  
Typically in such a configuration the colour flow is screened at larger distances and we deal  
with large size components of the Pomeron, so this contribution leads to a larger diffractive  
(dijet) cross section.  As a consequence the true value of the \lq\lq soft\rq\rq\ survival  
probability $S^2$ should be less than the estimate $S^2 \sim 0.1-0.15$ quoted  
above.  The conclusion from the CDF data is therefore consistent with the  
estimate $S^2 \sim 0.1$ at the Tevatron energy obtained from the soft  
rescattering model.  
  
It has recently been shown \cite{TI} that the $ep$ \cite{H1,ZEUS} and $p\bar{p}$ 
\cite{CDF} diffractive hard scattering data can be described in terms of the Soft Colour 
Interaction \cite{SCI} and Generalized Area Law \cite{GAL} models.  The unified 
description is obtained by implementing these models in the Monte Carlo generators LEPTO 
\cite{GI} for $ep$ and PYTHIA \cite{PY} for $p\bar{p}$.  However this \lq soft colour\rq\ 
approach leads to a somewhat flatter $\beta$ dependence than is observed for the CDF diffractive 
data, when the model is tuned to describe the deep inelastic $ep$ 
data\footnote{Note that an important part of $\omega=S^2 T^2$ gap survival factor is 
automatically included in PYTHIA, which accounts for the possibility of soft rescattering in 
the underlying event (the factor $S^2$) and the bremsstrahlung in the \lq hard\rq\ subprocess 
(part of the $T^2$ factor).  However the Soft Colour model does not account for 
bremsstrahlung from the two-gluon dipole Pomeron state. That is why $x^+$ (or $\beta$) 
dependence in \cite{TI} is flatter than that observed by CDF \cite{CDF}.}.  Our approach is 
different.  Here we emphasize, qualitatively, that   
the $p\bar{p}$ and $ep$ hard diffractive data have been presented in Ref.~\cite{CDF} in a   
way which demonstrates rather directly the role played by the survival probability factors   
$S^2$ and $T^2$ of (\ref{eq:a3}), and which allows physical insight into the interpretation of   
the diffractive data.  
  
In summary, we have shown that the data for hard scattering processes containing a rapidity  
gap in $p\bar{p}$ collisions at the Tevatron and the data for diffractive DIS collected in $ep$  
collisions at HERA, are compatible with each other.  The QCD factorization approach  
appears to lead to an order of magnitude discrepancy between the data sets \cite{CDF}.  
However the breakdown of factorization is naturally explained by the much smaller chance of  
the rapidity gap surviving in $p\bar{p}$ collisions as compared to $ep$ interactions.  Indeed  
the size, and $\beta$ dependence, of the suppression of diffractive dijet production seen at the  
Tevatron \cite{CDF} is just what is expected from the population of the rapidity gap by  
underlying soft interactions and from bremsstrahlung associated with the presence of the hard  
subprocess.  
  
\section*{Acknowledgements}  
  
VAK thanks the Leverhulme Trust for a Fellowship.  This work was also supported by the  
Royal Society, PPARC, the Russian Fund for Fundamental Research (98-02-17629) and the  
EU Framework TMR programme, contract FMRX-CT98-0194 (DG 12-MIHT).


\end{document}